\documentclass[fleqn,usenatbib]{mnras}

\usepackage{newtxtext,newtxmath}

\usepackage[T1]{fontenc}

\DeclareRobustCommand{\VAN}[3]{#2}
\let\VANthebibliography\thebibliography
\def\thebibliography{\DeclareRobustCommand{\VAN}[3]{##3}\VANthebibliography}

\usepackage{graphicx}	% Including figure files
\usepackage{amsmath}	% Advanced maths commands
\usepackage{supertabular}
\defcitealias{Weltevrede2006}{W06}
\defcitealias{Malofeev2018}{M18}

\title[Search for pulsars at 111 MHz]{Search for pulsars in an area with coordinates $3^h < \alpha < 4^h$ and $+21^o < \delta < +42^o$}

\author[S. A. Tyul'bashev et al.]{
S. A. Tyul'bashev,$^{1}$ \thanks{E-mail: serg@prao.ru (SAT)}
G. E. Tyul'basheva, $^{2}$
\\
$^{1}$ P.N. Lebedev Physical Institute of the Russian Academy of Sciences, Astro Space Center, Pushchino Radio Astronomy Observatory,\\
Radiotelescopnaya 1a, Moscow reg., Pushchino, 142290, Russia\\ 
$^{2}$ Institute of Mathematical Problems of Biology, Russian Academy of Sciences,
Branch of the Keldysh Institute of Applied Mathematics,\\ 
Pushchino, 142290 Russia\\
}

\date{May 01, 2023}

\pubyear{ , 2023}

\begin{document}
%\label{firstpage}
\maketitle

\begin{abstract}
On the Large Phased Array (LPA) of Lebedev Physics Institute (LPI), a search for pulsars outside the Galaxy plane was carried out in a 300 sq. deg area. The search with a sensitivity 5-10 times better than that of previously conducted surveys was at a frequency of 111 MHz. The search was carried out in the summed power spectra. With an accumulation equivalent to 100 hours of continuous observations for each point of the area, 5 known pulsars were detected with a signal-to-noise ratio (S/N) from 20 to 1300 in the first harmonic of the spectrum. Average profiles were obtained for the detected pulsars. Estimates of the peak and integral flux densities of the found pulsars are given for individual sessions and for the power spectra summarized over 5.5 years, obtained using the developed method based on measurements of the height of harmonics in the power spectrum. No new pulsars have been detected in the area. Apparently, when searching for pulsars in the area, we have approached the lower limit of the luminosity of the second pulsars. The completeness of the survey is at the level of 0.5~mJy.
\end{abstract}

%: J0034-0721, J0323+3944, J0528+2200, J0611+3016, J0826+2637, J1239+2453, J1921+2153, J2234+2114


%%%%%%%%%%%%%%%%%%%%%%%%%%%%%%%%%%%%%%%%%%%%%%%%%%

%%%%%%%%%%%%%%%%% BODY OF PAPER %%%%%%%%%%%%%%%%%%

\section{Introduction}

At the moment, there are more than 3300 pulsars in the ATNF catalog (https://www.atnf.csiro.au/research/pulsar/psrcat/) [1]. Most of them are classical radio pulsars that emit a pulse at each or almost every revolution of a neutron star. Their search, as a rule, was carried out with the using of power spectra, allowing to select periodic signals.

The celestial sphere has been viewed repeatedly in pulsar search surveys, and the detection of new pulsars during the next search is associated with the appearance of new sensitive antennas, new recorders with wide bands, and new processing methods. For example, the commissioning of the 500-meter FAST radio telescope in China and the launch of the pulsar search program led to the discovery of 201 pulsars [2]. At the moment, the search has been carried out only in a small part of the northern hemisphere, mainly in the plane of the Galaxy. By the end of the survey, we can expect the discovery of hundreds of new pulsars. The commissioning of the LOFAR meter-wavelength radio telescope, which is a spaced antenna array operating on broadband dipoles, made it possible to conduct a search at a frequency of 135 MHz and detect 53 new pulsars [3]. Incoherent accumulation of power spectra and the search for individual pulsar pulses made it possible to detect almost 90 pulsars in daily monitoring observations of the LPA (Large Phased Array) LPI (Lebedev Physics Institute) radio telescope operating at a frequency of 111 MHz ([4-6] and other papers).

A natural question arises about the meaning of the search for new pulsars, because only a small number of known pulsars have been studied in detail, and during the search, more and more weak objects are found, the full study of which is difficult. If these new weak pulsars are the same in their properties as the already known pulsars, then it is easier to study stronger objects.

The answer is that search, first of all, make it possible to conduct various kinds of statistical studies on pulsar samples, and also allow us to detect new classes of sources, which is impossible when studying individual sources. For example,
during the search for individual dispersed pulses, a new class of pulsars (rotating radio transient; RRAT) was discovered [7]. A study of randomly detected pulses with a duration of one minute showed that they belong to a white dwarf or pulsar with a period of $P$ = 18 min [8]. Quite recently
work has appeared on a new radio pulsar with a period of 75.88~c [9], lying behind the so–called "death line". This is the area on the dependence of the pulsar period on the derivative of the period ($P/Pdot$), where, according to theoretical assumptions, pulsar radiation should not occur.

The luminosities of pulsars can differ by six orders of magnitude (see ATNF), and it is still unclear whether the lower limit of the luminosity of pulsars has been found in the surveys. The boundary of the average luminosity is constantly shifting to the region of weak flux densities [2]. This means that the luminosity function changes, and it can tell us about the fundamentally observed number of pulsars in the Galaxy. In addition to these points, complete samples are the best for statistical research, i.e. samples containing all pulsars up to some flux density at a given frequency, or all pulsars up to some luminosity limit. Obtaining such homogeneous and complete samples is very difficult, taking into account the known strong variability of pulsars, both their own and related to the interstellar medium.

In [6], graphs are presented showing the expected sensitivity in the pulsar search survey in the PUshchino Multibeam Pulsar Search (PUMPS). According to these graphs, for pulsars with a dispersion measure $DM < 100-200$ pc/cm$^3$, with the accumulation of data over seven years, the sensitivity of the survey may be an order of magnitude higher than that of all surveys conducted in the world to search for pulsars. Earlier, using four-year observations obtained with low time-frequency resolution, we were able to detect 42 classical second pulsars (see https://bsa-analytics.prao.ru/pulsars/). In this paper, the area of 300 sq. deg. and processing 5.5 years of data with high time-frequency resolution is considered.

\section{Observations}

\begin{table*}
	\centering
	\caption{Known pulsars of the studied area}
	\label{tab:example_table}
	\begin{tabular}{lcccccccc}
	\hline
Name	& $P$,(s) &$DM$, & Publication & Dist,& RT & $\nu_{obs}; \Delta\nu; \Delta\nu_{chan}$ & $\Delta t$, & S/N 	\\
	    &     &pc/cm$^3$ &year     & kpc  &    & MHz  & min &  	\\
	\hline
0301+35  &0.568 & 57.4   &2016 [12]& 3.1  &Arecibo& 327;57;0.055& 1   & -;4.5\\
0302+2252&1.207 & 18.9   &2016 [11]&1.01 (2.74)&LPA&111;2.5;0.415& 3.5& 44;5.5\\
0323+3944&3.032 & 26.1   &1978 [13]&1.19 (21.7)&GB &400;16;2.0   & 2.4& $\simeq$1300;50.6\\
0342+27  &0.952 & 58.9   &2016 [12]& —    &Arecibo& 327;57;0.055& 1   & –;5.1 \\
0349+2340&2.420 & 62.9   &2019 [3] &3.74 (6.83)&LOFAR&135;32;0.012&60 &20;5.9\\
0355+28  &0.365 &48.6    &2020 [14]&1.78 (4.11)& GBT& 350;100;0.024& 2& 32;5.0\\
0358+4155&0.226 &46.3    &2014 [15]&1.48 (8.78)& GBT& 350;100;0.024& 2& 211;8.8\\
\hline
 \label{tab:tab1}
\end{tabular}
\end{table*}

The main instrument of the Pushchino Radio Astronomy Observatory (PRAO) is a LPA LPI, which is a meridian instrument. On the basis of one antenna field consisting of 16384 dipoles, two independent radio telescopes have been implemented at the moment [10]. One of these radio telescopes is used for monitoring round-the-clock sky survey in 96 spatial rays overlapping declinations of $-9^o < \delta < +42^o$, lined up along the meridian at with a distance between the beams of about half a degree.

Initially, the search for pulsars was carried out in data with low time-frequency resolution [11]. This data is recorded in 6 frequency channels, the channel width is 415 kHz, the sampling is 0.1 s. The data volume is approximately 1 terabyte per year, and they can be processed on household computers or weak servers. The search in data with low time-frequency resolution was carried out for pulsars with periods greater than 0.5~s. At $DM \simeq 30-40$ pc/cm$^3$, the sensitivity when searching for pulsars from these data begins to drop sharply due to dispersion smearing. No averaging of raw data was performed to search for pulsars with average profile widths greater than 100~ms, which leads to a deterioration in sensitivity for pulsars with wide ($>100$ ms) average profiles.

Since August 2014, parallel data recording with high time-frequency resolution has been carried out at the LPA LPI. The recording is conducted in 32 frequency channels with a width of 78 kHz and a sampling of 12.5~ms per point. The amount of data is about 35 terabytes per year. Data from August 2014 to the end of 2019 is stored on a disk shelf and is available for processing, data for 2020-2022 is stored on hard drives. The search in the data with high time-frequency resolution can be carried out for pulsars with periods $P > 0.025$~s. For pulsars with $P > 0.5$~c, the sensitivity begins to drop by $DM \simeq 100$ pc/cm$^3$ [6].

Processing of data with low time-frequency resolution showed that for some weak pulsars visible on the summed power spectra, it is not possible to obtain their average profile [6, 18]. To process and search for extremely weak pulsars in data with high time-frequency resolution, a special technique was created that allows extracting the main characteristics of the pulsar, their $P$ and $DM$ without obtaining an average profile. The data processing technique is described in detail in [6]. In this paper, the processing details are given to the extent necessary to understand the developed search methodology.

The developed pulsar search technique uses power spectra obtained using the Fast Fourier transform (FFT), and to increase sensitivity, these power spectra are added up over different days (different sessions). If the height of the harmonic in the spectrum is greater than the specified value in S/N units, then it is considered that a pulsar candidate has been detected. We get summed power spectra for each day, for each antenna beam by declination, for right ascensions in 1.5 time minute increments, for $DM$ sorted in the range of 0-1000 pc/cm$^3$, for possible pulse widths from 12.5 to 400 ms. Thus, for each point in the sky, we get from 1500 (survey mode) to 10000 (test mode for a candidate for a new pulsar) power spectra. To reflect the contents of the summed spectra, we build a $P/DM$ map, on which we mark with circles of different sizes all signals (harmonic heights of power spectra) having a signal-to-noise ratio greater than the specified one. By default, the minimum value of S/N=4 is set on the map, harmonics having S/N<4 are not shown on it. The S/N border can be shifted upwards, removing weak periodic signals from the map. If a pulsar hits the point under study, it is reflected on the map in the form of segments stretched along the $DM$. The second, third and other harmonics of the pulsar look like shorter segments. Their periods are $P/2$, $P/3$ and so on. Due to the low-frequency noise associated with radio sources scintillating on the interplanetary plasma and on the ionosphere, the baseline is not always successfully subtracted from the power spectrum. Low-frequency noises remain in it, so the search for pulsars by power spectra can be carried out only up to periods of 3-4 s. Summed periodograms are used to search for pulsars having $P>3$~c.

Physically, observations are recorded on two recorders, each of which has 48 LPA beams connected. For our work, we selected data from a recorder recording declinations of $+21^o < \delta < +42^o$. At these declinations, there is a minimum of interference, and the best sensitivity is achieved. Processing was carried out on a household computer, so the processing speed is low. For the test of a new search program working with data recorded with high time-frequency resolution, an average for all its characteristics area with $3^h < \alpha < 4^h$ coordinates was selected. The size of this area is equal to 1/48 of the total size of the viewing area. Processing of observations for 5.5 years lasted 1.5 months.

In the survey area at the beginning of August 2022, ATNF showed 199 pulsars having $P>25$~ms and $DM < 200$ pc/cm$^3$. The maximum number of known pulsars falls on the regions $5^h < \alpha < 6^h$ (9 pulsars) and $19^h < \alpha < 21^h$ (79 pulsars), corresponding to the directions to the anticenter and the center of the Galaxy. In four of the 24 areas in the catalog there is only one pulsar. The median number of pulsars per hour of right ascension is six. In our chosen area in the ATNF catalog there are six known second radio pulsars. In addition, on the website http://www.naic.edu /$\sim$deneva/drift-search/ [12] the source J0342+27 was found, which did not get to the ATNF catalog. Thus, at the moment there are seven known pulsars in the study area.

\section{Detection of known pulsars} 

\begin{figure*}
\begin{center}
	\includegraphics[width=\textwidth]{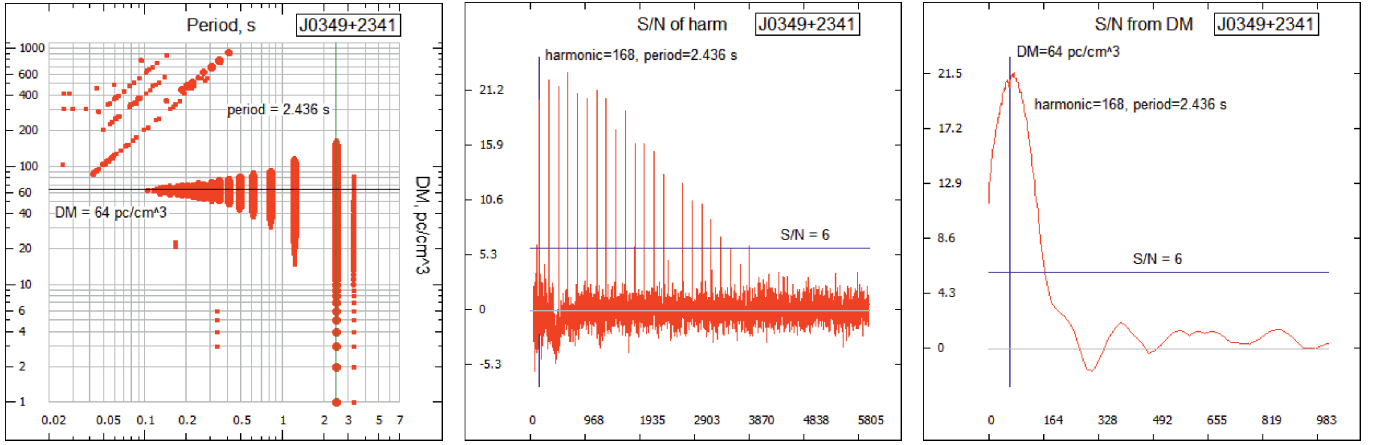}
    \caption{Pulsar J0349+2341. Left part of the picture: FFT-map $P/DM$. The circles show the harmonics of the summed power spectra with a $S/N \ge 6$. Horizontally it is the harmonic period in seconds, vertically it is $DM$ in pc/cm$^3$. The middle part of the figure: the summed power spectrum for 1800 days obtained for the assumed $DM$=64~pc/cm$^3$. Horizontally it is the number of the point in the power spectrum, vertically it is S/N. The right part of the figure shows the height of the first harmonic of the summed power spectrum on the $DM$. Horizontally it is $DM$ in pc/cm$^3$, vertically it is S/N.}
    \label{fig:fig1}
\end{center}
\end{figure*}

A search program using the summation of power spectra for the entire observation period has detected five of the seven known pulsars. Information on them is given in Table~1. Columns of the table: "name" – the name of the pulsar in the J2000 annotation; "$P$" – the period; "$DM$" – the dispersion measure, "year of publication" – the year of publication of the paper with the detected pulsar and the link to the paper; "Dist" – the distance to the pulsar; "RT" – the name of the radio telescope; "$\nu_{obs}; \Delta\nu; \Delta\nu_{chan}$" – the characteristics of the pulsar in the surveys: the central frequency of observations, the full reception band,
the width of the frequency channel; "$\Delta t$" – the accumulation time in one observation session according to the survey data; "S/N" – the estimate in S/N units of the height of the first harmonic in the summed power spectra and in the best observation session according to the LPA data.

For pulsars J0301+35 and J0342+27, it was not possible to find papers that would talk about their first detection. These pulsars are only available on the website where the survey of AO327 (http://www.naic.edu/$\sim$deneva/drift-search/) is presented. For these pulsars, the table provides a link to an paper describing the survey. For the pulsar J0349+2340 is the first independent confirmation after its detection. The "Dist" column indicates the distance to the pulsar according to the ATNF catalog, and an estimate of the distance to the pulsar is given in parentheses, obtained from the value of the S/N accumulated in the total power spectra in observations at the LPA LPI, assuming that the observed harmonic height decreases proportionally to the square of the distance to the pulsar. The "RT" column contains the names of the telescopes: GBT – Green-Bank full-rotation 100-meter telescope; LPA – meridian antenna array $200 \times 400$~m with full aperture; GB – Green-Bank meridian 90-meter telescope; LOFAR is a distributed antenna array with a synthesized beam; Arecibo is a meridian 300–meter telescope.

\begin{figure*}
\begin{center}
	\includegraphics[width=\textwidth]{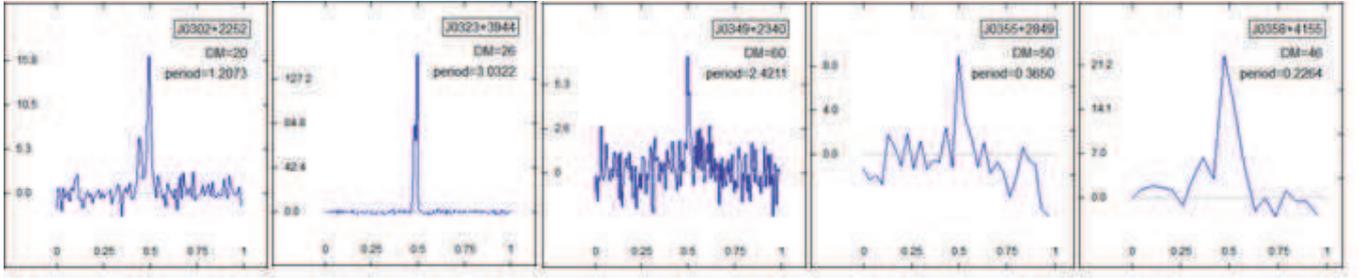}
    \caption{Average pulsar profiles from Table 1, generated by the pulsar processing program. Vertically it is the S/N of the profile, horizontally it is the phase. The maximum in the profile (pulse phase) after the addition of pulses can occur at any point in the profile, but in the figures the maxima are shifted to the 0.5 phase.}
%    \label{fig:fig2}
\end{center}
\end{figure*}

Summation of power spectra for all signals was performed on average for 1800 days (3.5 min per sessions). Taking into account the time of passage of the source through the meridian, the total time of signal accumulation at each point is approximately one hundred hours.

Separately, we note the sources J0301+35 and J0355+28, which apparently have typos of periods in ATNF. On the Arecibo (http://www.naic.edu/$\sim$deneva/drift-search/) survey site, a period of 0.568~s is indicated for the pulsar J0301+35, and a period of 0.146~s is indicated in ATNF for a pulsar with the same name and $DM$. A similar situation with the pulsar J0355+28 is on the website accumulating pulsars (http://astro.phys.wvu.edu/GBNCC /), detected in GBT, the period is 0.365 c, and on the ATNF website – 0.094~s.

As an example, here are the figures generated by the processing program for the pulsar J0349+2340. This pulsar is the weakest pulsar we have found in the study area. Fig.~1 shows 22 harmonics on the map and on the total power spectrum.

Pulsars J0301+35 and J0342+27 were not detected in the search by summing power spectra. Neither of these pulsars were detected in the LOFAR survey [3], which has a sensitivity of 2-3~mJy [6] when converted to 111~MHz. If pulsars are flashy, or have strong variability, or they have a large nulling part, then summing the power spectra for all days will not be the optimal way to search for them. We conducted a search for the emission of these pulsars in the power spectra calculated for individual days, assuming that over 5.5 years of observations there may be days when these pulsars were strong enough to be seen in separate sessions.

For pulsars J0301+35 and J0342+27, there were 3-5 "suspicious" days out of almost 2000 checked. Profiles with pulses (S/N=4.5–5) were obtained, in which $P$ and $DM$ are close to the catalog ones. However, if the pulses are real, they should be detected in the average profiles at close periods and $DMs$. In addition to this, the pulse duration cannot be less than the amount of dispersion smearing in the frequency channel on a known pulsar $DM$. The pulses we found are detected only at one $P$ and $DM$, so we believe that the detected signals do not belong to pulsars J0301+35 and J0342+27. Below we give upper estimates of the flux density for these pulsars, based on the height of peak signals estimated on "suspicious" days.

Fig.~2 shows the average profiles of the pulsars detected by us (see Table~1) for the days with the most characteristic, from our point of view, profiles. In the obtained average profiles, the program estimates the height of the pulse in units of S/N. Using the value of S/N, we can obtain an estimate of the peak flux density ($S_p$) under the assumption that we know the magnitude of the rms deviations of the noise track ($\sigma_n$ is the fluctuation sensitivity corresponding to S/N=1) in the direction of the pulsar in units of flux density.

According to the isophotes in [16] obtained at a frequency of 178 MHz close to the central frequency of LPA observations, the difference in background temperatures on the entire studied area does not exceed 10\%, so background noise will be almost the same throughout the area. According to the radiometric gain formula, it is possible to estimate $\sigma_n$ over the entire area, and, knowing the S/N of the pulse, get the $S_p$ of the pulse in Jy. To obtain the integral flux density ($S_i$), it is necessary to sum up the points of the average profile taking into account the known value of $S_p$.

The peak flux density, obtained in units of S/N and visible in Fig.~2, is the value estimated from the data on the average profile. However, in observations on the antenna array, we observe only a part of the emitted energy. To obtain an estimate of the actual flux density, it is necessary to make corrections that take into account the antenna features: the zenith distance of the source; the mismatch of the coordinates of the LPA beam and the pulsar; the correction for the envelope of the formed eights of the LPA beams. All these amendments are clearly shown in Fig.~1 in [10], where it is seen that the real flux densities can be several times greater than the observed fluxes.

Thus, for records in which the average profile is visible after processing, $S_p$ and $S_i$ can be estimated if the fluctuation sensitivity is known. These best recordings were selected according to power spectra for individual days, where harmonics with the highest S/N were observed. If you link the harmonic height for a given date to a known flux density defined for that date, you can estimate the flux densities for any dates by measuring the harmonic height in the power spectrum in units of S/N and assuming that the fluctuation sensitivity varies slightly from day to day. That is, it is possible to recalculate the flux density by measuring the harmonic height in the summed power spectrum. Of course, these rough estimates will apply to pulsars whose own (internal) or external (due to the interstellar medium) variability does not significantly contribute to the growth of harmonics when adding spectra for different dates. For such pulsars, the harmonic height in the total power spectrum should grow as the square root of the number of stacked sessions.

As shown in [18], for part of the observation sessions, the quality of the noise track is so low that these days are removed from processing. For the remaining sessions, the noise track may change from day to day, due to weather conditions, the physical condition of the antenna and other reasons. These factors can worsen the final ratio of the S/N in the summed power spectrum by 20-30\%.

Thus, the system of estimates of flux densities in this work is tied to the assessment of the fluctuation sensitivity of the LPA LPI in the studied area. Our estimate $\sigma_n = 0.34$~Jy was obtained assuming a sampling of 12.5 ms, a reception band of 2.5 MHz, an effective antenna area in the zenith direction of 45,000 sq. m., a system temperature of 1000K and one received polarization.

Finally, the estimate of the peak flux density in the average profile is estimated in S/N units, taking into account the number of stacked pulses and corrections for antenna features. The integral flux density in a single session is estimated by summing all the points of the average profile after subtracting the baseline. It is assumed that the height of the first harmonic of the power spectrum in the observation session uniquely reflects the height of the average profile in the same session. It is assumed that the S/N of the first harmonic in the summed power spectrum, taking into account all the losses described above, is associated with the S/N of the harmonic in a single session.

Pulsars are objects with variable flux density. There are days when the observed pulsar flux density may be higher than average. The average profiles in Fig.~2 are based on such "best" days selected from 2000 observation sessions. Therefore, the extracted estimates of the flux density for individual sessions should be overestimated compared to the average fluxes of $S_p$ and $S_i$.

For the summed power spectra and power spectra in individual sessions, the height of the first harmonic was estimated in S/N units (column 9 in Table~1). After summation, the S/N increases in comparison with one session. According to our estimates, the S/N of harmonics in the summed power spectra has increased 36 times over 5.5 years (for details, see [18]). The typical (average) S/N in the power spectrum in one observation session for the weakest of the confirmed pulsar (J0349+2340) should be 20/36 = 0.56. In the same way, estimates can be made for all pulsars. We can associate the value of the S/N in the summed spectrum with the value of the S/N in a single observation session. Below we have expressed estimates of peak ($S_p$) and integral ($S_i$) flux densities in conditional formulas:\\
$S_{p-one-sess}=\sigma_n \times [S/N_{peak-av-prof-in-one-sess}/N^{1/2}]\times k$,\\
$S_{i-one-sess}=\sigma_n \times [S/N_{int-av-prof-in-one-sess}/N^{1/2}]\times k,$\\
$S_{p-sum-all-sess}=S_{p-one-sess}\times \frac{(S/N_{first-harm-sp-sum})/36}{S/N_{first-harm-one-sess}}$,\\
$S_{i-sum-all-sess}=S_{i-one-sess}\times \frac{(S/N_{first-harm-sp-sum})/36}{S/N_{first-harm-one-sess}}$,\\
where $N$ is the number of periods in one session for a given pulsar, and $k$ is its antenna corrections.

In the Table~2 estimates of pulsar flux density are given. The name of the pulsar is given in the first column. Columns 2 and 3 show the peak flux densities for observations in one session ($S_{p1}$) and when averaging over the entire period ($S_{p1-av}$). Columns 4 and 5 show the integral flux densities for observations in one session ($S_{i1}$) and during averaging ($S_{i 1-av}$). Columns 6 and 7 show an estimate of the flux density from the papers of other authors ($S_{i2}$), a reference to the original work and the frequency ($\nu$ in MHz) at which the estimate was obtained. Column 8 shows the expected integral flux density ($S_{i3}$) at a frequency of 111 MHz, recalculated from $S_{i2}$ under the assumption that the spectral index is 1.7. The integral flux densities before recalculation were taken from two papers where observations of pulsars on LOFAR are given. For pulsars J0301+35 and J0342+27, the upper estimates of peak ($S_{p1-av}$) and integral ($S_{i1}$, $S_{i1-av}$) flux densities are given under the assumption that the width of the pulse profile at a frequency of 111 MHz is the same as on the profiles in the paper with the detection of these pulsars [12]. The upper estimates of the flux density of pulsars J0301+35 and J0342+27 are conservative estimates obtained from peaks observed in the noise of the average profiles of these pulsars. The actual flux densities seem to be 1.5 or more times less if these pulsars actually have radiation at a frequency of 111 MHz.

The estimates of the flux densities in Table~2 show that for all pulsars, the integral flux densities estimated on days with observed maximum S/N in the average profile and power spectrum exceed the expected values of the integral flux density according to LOFAR observations. Estimates of the average integral flux density can serve as a lower estimate of the integral flux density. For all pulsars, with the exception of J0358+4155, the lower estimates of the integral flux density are from 4 to 8 times less than the expected flux densities obtained by the LOFAR. For J0358+4155, the estimates of the integral flux density are several times higher than the expected estimates. That is, the proposed method of estimating the flux density can be used if there is no other reliable way to get an estimate of the flux.

\section{Candidates for new pulsars and their verification}

\begin{table*}
	%\centering
	\caption{Known pulsars of the studied area}
	\label{tab:example_tab}
	\begin{tabular}{lccccccc}
	\hline
Name	& $S_{p1}$&$S_{p1-av}$& $S_{i1}$&$S_{i1-av}$&$S_{i2}$&$\nu$&$S_{i3}$\\
	\hline
0301+35  & $<96$  & $<2.7$ & $<2$ & $<0.06$ & - & - & - \\
0302+2252& 1636   & 122    & 54   & 4.0 &20 [17]  & 129& 26\\
0323+3944& 10584  & 2517   & 168  & 40  &109.6 [3]& 135&153\\
0342+27  &$<150$  &$<4.2$  &$<8$  &$<0.24$&  -  & — & — \\
0349+2340& 500    & 10.0   & 27   & 0.5 & 3 [17]& 129& 3.9\\
0355+28  & 250    & 25     & 15   & 1.5 &4.5 [3]& 135& 6.3\\
0358+4155& 1400   & 930    & 152  & 101 &21.9 [3]& 135& 31\\
 \hline
 \label{tab:tab2}
\end{tabular}
\end{table*}

\begin{figure*}
\begin{center}
    \includegraphics[width=\textwidth]{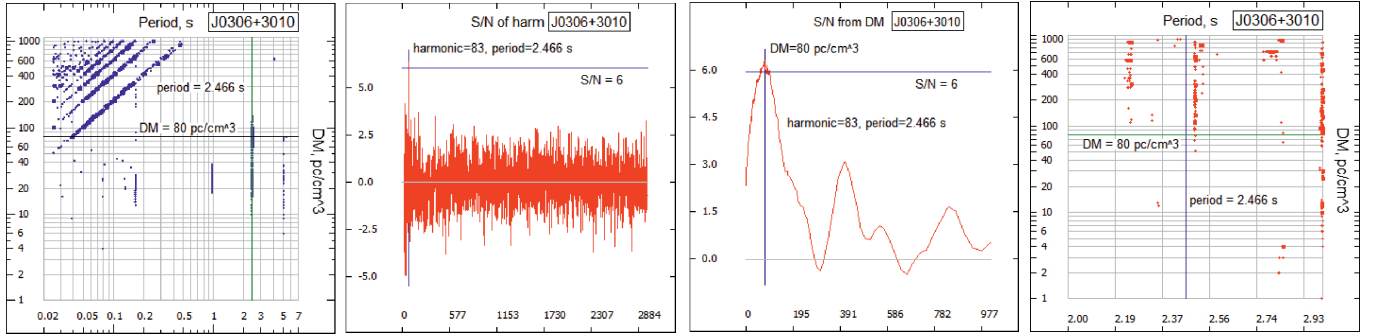}
    \caption{Pulsar candidate J0306+3010 ($DM = 80$~pc/cm$^3$, $P = 2.466$~s). The first three panels are: the FFT map $P/DM$, the summed power spectrum for $DM = 80$~pc/cm$^3$, and the dependence of S/N on $DM$ for $P = 2.466$~s. The map shows harmonics with a $S/N \ge 4$, the rest is similar to Fig.~1. The rightmost figure is the FFA map $P/DM$ (map of summed periodograms). The intersection of the vertical ($P$) and horizontal ($DM$) lines shows the place on the map where we expected the pulsar candidate found on the FFT map to appear. All drawings are generated by the processing program.}
    \label{fig:fig3}
\end{center}
\end{figure*}

In [4,18] a set of criteria was developed that a new pulsar found when summing power spectra should satisfy: the height of the harmonic in the power spectrum is $S/N > 6-7$; at least two harmonics are observed in the spectrum; there are several independent individual records according to which the average profile is constructed; the pulsar is detected in one or two adjacent antenna beams; the pulse height in the average profile has a clear dependence on the $DM$ being tested.

It was shown in [6] that for the weakest pulsars there is often not a single record from which an average pulsar profile can be obtained. At the same time, the pulsars themselves, known from observations on other telescopes, are reliably determined in the summed power spectra.

We conducted a search for pulsars in our chosen area: $3^h < \alpha < 4^h$ and $+21^o < \delta < +42^o$. No candidates with two harmonics in the power spectrum or the height of the first harmonic $S/N \ge 7$ were found. Two pulsar candidates with one harmonic in the spectrum have been found
power $6 < S/N < 7$. The periods of these candidates $1.6 < P < 2.5$~s are close to the region where low-frequency noise is observed in the power spectra ($P > 3-4$~s, [6]). Other signs of the pulsar are also confirmed: the detection of a candidate in only one beam of the LPA LPI and a clearly expressed dependence of the height of the first harmonic on the $DM$ being tested.

The period of the observed harmonic on the $P/DM$ map is close to the area where interference caused by poor baseline subtraction is regularly observed, and this calls for additional candidate verification. Generally speaking, the problem of removing low-frequency noise from the power spectra is a common problem that arises when searching for pulsars, regardless of the characteristics of the survey, the central frequency of the observations, the number of frequency channels, the sampling time of the raw data point, etc. (see, for example, paragraph 6.1.3.3. in the handbook of pulsar astronomy [19]). The appearance of low-frequency noise is due to the fact that the received noise has low-frequency components that are associated with the background of the Galaxy and interference that cannot be completely removed. For the LPA LPI radio telescope, the problem of the appearance of low-frequency noise in the power spectra is aggravated by the fact that the telescope has a high fluctuation sensitivity and a low angular resolution of approximately $0.5^o \times 1^o$. Therefore, in the recordings, among other things, the effects of confusion of extended (having angular dimensions $\theta > 1^{\prime \prime}$ ) and compact (scintillating) radio sources (with angular dimensions $\theta < 1^{\prime \prime}$) are observed simultaneously. In addition, ionospheric scintillating with characteristic time scales from several to several tens of time seconds contribute to the level of low-frequency noise [20]. Subtracting the background of the Galaxy from the raw data, taking into account two effects of confusion and various kinds of interference, leads to the appearance of low-frequency noise, lying mainly at frequencies corresponding to periods greater than 3-4~s [6].

The candidates we found have periods of less than 3 s, but they have only one harmonic, and at the location of the second harmonic in the summed power spectra there are no signals having $S/N > 3$. One harmonic in the power spectrum can be observed in coaxial rotators, or pulsars having such a large $DM$ that pulses pulsars are smeared for the entire period due to dispersion smearing in frequency channels or due to pulse scattering. A preliminary estimate of the $DM$ of the candidates (for first $DM \sim 80$~pc/cm$^3$, and for the second $DM \sim 30$~pc/cm$^3$) shows that dispersion smearing and
scattering slightly widen the pulses.

There are a few pulsars of coaxial rotators. For example, if the ATNF catalog displays the period and the half-width of the average profile in the table, then you can get a picture at the output that shows that the typical half-width of the pulse is several percent of the period, whereas for coaxial rotators this ratio should be close to 50\%. Out of about 1,500 pulsars with a measured pulse half-width, there are about ten possible coaxial rotators. Therefore, there is no reason to assume that the candidates we found are coaxial rotators. Additional evidence is needed to confirm that new pulsars have been found.

The appearance of harmonics can be caused by low-frequency noise inherent in the power spectra obtained using FFT. However, the search can also be carried out using periodograms (Fast Folding Algorithm – FFA), whose noise in the region of large periods is low, and the sensitivity is higher than when searching using FFT [21]. We conducted a search for candidates using periodograms, but it was not possible to confirm that the candidates are pulsars. To illustrate, we present Fig.~3, showing the pulsar candidate we found on FFT maps and the absence of this candidate in FFA maps.

\section{ANALYSIS OF THE RESULTS}

It was noted in [6] that 14\% of known pulsars (13 out of 95) with $DM < 100$~pc/cm$^3$ were not detected in observations at the LPA LPI during the search. In the section "Detection of known pulsars" it was shown that two out of seven known pulsars were not detected (28\%), despite the high sensitivity that is realized when summing power spectra (see Table~2). Similar "non-detection" of known pulsars is noted in other surveys [22, 23]. Special consideration of the reasons why known pulsars may not be visible during search operations, by the authors no original paper is being done. The most likely reasons for non-detection may be a cut off in the spectrum or radiation features.

The sensitivity limit of the AO327 survey [12] according to Fig.~4 of the paper [6] is approximately 2.5~mJy, which corresponds to $S/N = 6$ at a frequency of 111~MHz. Considering that the AO327 survey was made at a frequency of 327~MHz and assuming that the pulsar was not detected due to the cut off of the spectrum that began at a frequency of 327~MHz, the spectral index should be equal to 3 ($S\sim \nu^{-\alpha}$). Since the pulsars J0301+35 and J0342+27 were detected in the survey [12] at $S/N \sim 10-20$, their spectral index in the region of the cut off of the spectrum should be even greater than 3.

The search was carried out using summed power spectra. If a large number of nulling is observed in the meter range or there are long periods of switching off the pulsar radiation, the signal accumulation will be ineffective [19]. A similar situation may occur in the case of strong variability of pulsar radiation.

In total, five known pulsars were detected in the area. These pulsars are located at distances from 1 to 3.7 kpc and have $DM$ from 18.9 to 62.9 pc/cm$^3$. As shown in column 5 of Table 1, the detected pulsars can be detected from distances that are 2-3 times or more than the distances at which they are actually located. We will estimate the maximum possible $DM$ (up to he edge of the halo) in the direction of these pulsars according to the YMW2016 model [24]: J0302+2252 (50 pc/cm$^3$), J0323+3944 (118 pc/cm$^3$),J0349+2340 (77 pc/cm$^3$), J0355+28 (97 pc/cm$^3$), J0358+4155 (175 pc/cm$^3$). For hypothetical pulsars located in the direction of J0323+3944 and J0358+4155, the sensitivity of observations is sufficient to detect them to the edge of the halo, taking into account the increasing $DM$. If possible new pulsars have the same distribution in the sky and the same properties ($P; DM$) as the known pulsars in the studied area, then in the survey we would have to open 10-20 new ones for each known pulsar (the volume grows like a cube of distance, therefore increasing the distance at which the pulsar is visible, 2 times leads to an increase in volume by $2^3 = 8$ times).

According to [6], the sensitivity when processing survey data recorded in 32-channel mode with daily observations for 5.5 years can reach 0.1~mJy for the integral flux density. Taking into account the peculiarities of the radiation pattern and the distance between the beams of the LPA, the completeness of the survey for second pulsars in [6] is estimated as 0.5~mJy. No new pulsars were detected in the conducted search, which means that they have been exhausted to the level of at least 0.5~mJy at a frequency of 111~MHz in the studied area.

As already mentioned in the Introduction, the maximum distribution of the luminosity function shifts to the region of increasingly lower luminosities with increasing search sensitivity [2]. The area we studied is small enough to make strict statistical conclusions, but the absence of new pulsar detections suggests, apparently, that we have approached the lower limit of the luminosity of second pulsars.

\section{Conclusions}

The search for pulsars with periods from 0.025~s and $DM$ up to 1000 pc/cm$^3$ was carried out using data with high time-frequency resolution. The achieved sensitivity in the survey was 5-10 times higher than in all previous surveys. Extremely weak pulsars in the survey can be detected up to 0.1~mJy, and the completeness of the survey is guaranteed at 0.5~mJy at a frequency of 111~MHz. The survey was conducted in the direction close to the direction of the anticenter, in the area $3^h < \alpha < 4^h$ and $+21^o < \delta < +42^o$. No new pulsars have been detected, which indicates their exhaustion in a volume that is 10-20 times higher than the previously covered volumes in the Galaxy.

Five known second pulsars located in the investigated area were detected. For the pulsar J0349+2340 is the first independent confirmation. Three out of five pulsars have not been detected on the LPA before. Average pulsar profiles were obtained. Estimates of peak and integral flux densities of detected pulsars are given. They are obtained using a specially developed method that uses the estimation of the height of harmonics in the summed power spectrum and in the power spectra extracted in individual sessions. Two known pulsars were not detected during the search. Most likely, these are pulsars with nulling or intermittent pulsars.

\section*{Acknowledgements}
The study was carried out at the expense of a grant Russian Science Foundation 22-12-00236, https:// rscf.ru/ project/ 22- 12- 00236/.

\section*{Bibliography}
1. R. N. Manchester, G. B. Hobbs, A. Teoh, and M. Hobbs, Astron. J. {\bf 129}, 1993 (2005).

2. J. L. Han, Chen Wang, P. F. Wang, et al., Research in Astronomy and Astrophysics {\bf 21}, 107 (2021).

3. S. Sanidas, S. Cooper, C. G. Bassa, et al., Astron. and Astrophys. {\bf 626}, A104 (2019).

4. S. A. Tyul’bashev, V. S. Tyul’bashev, M.A. Kitaeva, et al., Astronomy Reports {\bf 61}, 848 (2017).

5. S. A. Tyul’bashev, V. S. Tyul’bashev, and V. M. Malofeev, Astron. and Astrophys. {\bf 618}, A70 (2018b).

6. S. A. Tyul’bashev, M. A. Kitaeva, and G. E. Tyul’basheva, arXiv:2203.15540 (2022).

7. M. A. McLaughlin, A. G. Lyne, D. R. Lorimer, et al., Nature {\bf 439}, 817 (2006).

8. N. Hurley-Walker, X. Zhang, A. Bahramian, et al., Nature {\bf 601}, 526 (2022).

9. M. Caleb, I. Heywood, K. Rajwade, et al., Nature
Astronomy {\bf 6}, 828 (2022).

10. V. I. Shishov, I. V. Chashei, V. V. Oreshko, et al., Astronomy Reports {\bf 60}, 1067 (2016).

11. S. A. Tyul’bashev, V. S. Tyul’bashev, V. V. Oreshko, and S. V. Logvinenko, Astronomy Reports {\bf 60}, 220 (2016).

12. J. S. Deneva, K. Stovall, M. A. McLaughlin, et al., Astrophys. J. {\bf 821}, 10 (2016).

13. M. Damashek, J. H. Taylor, and R. A. Hulse, Astrophys. J. {\bf 225}, L31 (1978).

14. A. E. McEwen, R. Spiewak, J. K. Swiggum, et al., Astrophys. J. {\bf 892(2)}, 76 (2020).

15. K. Stovall, R. S. Lynch, S. M. Ransom, et al., Astrophys. J. {\bf 791}, 67 (2014).

16. A. J. Turtle and J. E. Baldwin, Monthly Not. Roy. Astron. Soc. {\bf 124}, 459 (1962).

17. D. Michilli, C. Bassa, S. Cooper, et al., Monthly Not. Roy. Astron. Soc. {\bf 491(1)}, 725 (2020).

18. S. A. Tyul’bashev, M. A. Kitaeva, V. S. Tyul’bashev, et al., Astronomy Reports, {\bf 64}, 526 (2020).

19. D. R. Lorimer and M. Kramer, Handbook of Pulsar
Astronomy (Cambridge, UK: Cambridge University
Press, 2012).

20. I. V. Chashei, S. A. Tyul’bashev, V. I. Shishov, et al., Space Weather {\bf 14}, 682 (2016).

21. V. Morello, E. D. Barr, B. W. Stappers, E. F. Keane, and A. G. Lyne, Monthly Not. Roy. Astron. Soc. {\bf 497}, 4654 (2020).

22. R. N. Manchester, et al., Monthly Not. Roy. Astron. Soc. 279, 1235 (1996).

23. J. S. Deneva, K. Stovall, M. A. McLaughlin, S. D. Bates, P. C. C. Freire, J. G. Martinez, F. Jenet, and M. Bagchi, Astrophys. J. 775, 51 (2013).

24. J. M. Yao, R. N. Manchester, and N. Wang, Astrophys. J. 835, 29 (2017).

\bibliographystyle{mnras}
%\bibliography{serg1}

\end{document}